\documentstyle[prd,aps,twocolumn,amstex,latexsym,epsfig,amsfonts]{revtex}

\newcommand{\inner}[2]{\left(#1|#2\right)}

\newcommand{\nn}{\nonumber\\} 
 
\newcommand{\f}[1]{\mbox{\boldmath$#1$}}

\newcommand{\vau}{\mbox{\boldmath$v$}}
\newcommand{\na}{\mbox{\boldmath$\nabla$}}
\newcommand{\bea}{\begin{eqnarray}}
\newcommand{\ea}{\end{eqnarray}}
\newcommand{\eea}{\end{eqnarray}}

\begin{document} 

\wideabs{
\title{Gravity wave analogs of black holes}
\author{Ralf Sch\"utzhold and William G.~Unruh}
\address{ 
Department of Physics and Astronomy,
University of British Columbia,
Vancouver, B.C., V6T 1Z1 Canada\\
Electronic addresses: {\tt schuetz@@physics.ubc.ca, unruh@@physics.ubc.ca}
}
\date{\today}
\maketitle
\begin{abstract} 
It is demonstrated that gravity waves of a flowing fluid in a shallow basin 
can be used to simulate phenomena around black holes in the laboratory. 
Since the speed of the gravity waves as well as their high-wavenumber
dispersion (subluminal vs.\ superluminal) can be adjusted easily by varying
the height of the fluid (and its surface tension) this scenario has certain 
advantages over the sonic and dielectric black hole analogs, for example, 
although its use in testing quantum effects is dubious.
It can be used to investigate the various classical instabilities associated 
with black (and white) holes experimentally, including positive and negative 
norm mode mixing at horizons. 
\end{abstract}    
}

\section{Introduction}\label{intro}

One of the most fascinating predictions of Einstein's theory of general 
relativity is the potential existence of black holes -- i.e.\ space-time 
regions from which nothing is able to escape.
Perhaps no less interesting are their antonyms: white holes
(nothing can penetrate).
Both are described by solutions of the Einstein equations and are related to 
each other via time-inversion, see e.g.\ \cite{misner}.
 
As it is well-known, these objects feature many novel  properties: 
For example, for orbits sufficiently close to the horizon (i.e.\ for $r<3M$) 
one observes \cite{centrifugal} an inversion of the centrifugal acceleration.

Rotating black holes as described by the Kerr metric admit 
unstable modes under certain conditions, i.e.\ solutions of the wave equation 
growing in time without any bound, 
see e.g.\ \cite{kerrunstable,bomb}. 
This phenomenon is related to the mechanism of superradiance 
\cite{super} which allows 
one to extract energy from the rotation of the Kerr black hole, 
cf.\ \cite{misner}.
 
White holes are unstable  \cite{eardley} to exponential build-up of 
energy on the white hole Cauchy horizon on the classical level, 
as well as on the quantum level \cite{zeldovich,rama,antievapor}.

The presence of both Cauchy and particle horizons 
(white and black hole horizons), such as in the interior of
a Reissner-Nordstr\"om metric, can have further instabilities, 
see e.g.\ \cite{inner}.
 
Another striking effect is the evaporation \cite{hawking} of  
black holes due to quantum  effects. 
This observation can be interpreted as a confirmation of their thermodynamical 
interpretation \cite{bekenstein} relating purely 
geometrical quantities, such as surface gravity and surface area, to thermal 
properties, such as temperature and entropy.
 
Fortunately it seems unlikely that one can observe black holes in the 
laboratory (see, however, e.g.\ \cite{lhc}).
Analogs, which obey similar equations of motion to fields around a black hole
raise the possibility of demonstrating some of the most unusual properties of
black holes in the laboratory.
This is the basic idea of the black and white hole analogs (Dumb holes) 
originally proposed by Unruh  in Ref.\ \cite{unruh}.
The sonic analogs established there are based on the observation
that sound waves in flowing fluids are (under appropriate conditions)
governed by the same wave equation as a scalar field in a curved space-time.
The acoustic horizon, which occurs if the velocity of the fluid exceeds the
speed of sound within the liquid, acts on sound waves exactly as a black hole
horizon does on, for example, scalar waves.
 
After the original proposal in Ref.\ \cite{unruh} the sonic analogs have 
been the subject of several investigations, see e.g.\
\cite{visser,volovik,garay}.
Although the kinematics of the waves propagating within the black and 
white hole analogs are governed by the same equation as those in a 
curved space-time, the dynamics of the effective metric itself are not 
described by the same laws as gravity (i.e.\ the Einstein equations)
in general\footnote{
There are, however, possibilities to reproduce the Einstein equations
even in non-gravitational systems, see \cite{emergent}.}.

In this way the analogs allow one to  separate  the dynamical  
effects of gravity (following from the Einstein equations) from more 
general (kinematic) phenomena, cf.\ \cite{without,rosu}.
 
In addition to the sonic analogs there exist proposals for black or
white hole analogs based on the propagation of light in dielectric media 
(instead of sound), see e.g.\ \cite{rosu,novello,leo,reznik,dielectric}, 
and of other sorts of waves in for example liquid Helium 3, 
see e.g.\ \cite{volovik}.
These scenarios avoid some of the difficulties associated to the sonic 
analogs but can have other problems.
 
The challenge in making such analogs to horizons is in preparing a medium in
which the waves are stopped from propagating out from some region. In the
analogs where the flow of a medium is used to drag the waves at a velocity
corresponding to the velocity of the waves, one requires a sufficiently low
velocity that the experiment could be contemplated. 
The speed of sound depends on the equation of state $p=p(\varrho)$ 
only and therefore is hard to  adjust by external parameters.
In the case of the analogs based on light, the velocity of the 
(quasi) photons is determined by the effective permittivity and 
permeability of the medium, which can also be hard to manipulate, 
and especially hard to make a sufficiently low group {\em and} phase 
velocity of light (cf.\ \cite{leo,dielectric,pessi}).
 
Consequently we were led to look for another kind of waves traveling at a 
velocity that can be controlled more easily.
One promising candidate is gravity waves (surface waves) in a shallow basin 
filled with a liquid.
As we shall see in the following, long gravity waves within a flowing
fluid are {\em also} governed by the same wave equation as a scalar field 
in a curved space-time. 
In addition, the speed of the long gravity waves can be adjusted very 
easily by varying the depth of the basin. 
Furthermore the fluid flow in such a basin is easily manipulated. 
Because of the low velocity of these waves, quantum
effects would not be observable, but many of the classical features of black
holes (including the positive and negative norm mixing at the horizon which is
closely related to the quantum evaporation effects) could be investigated.
Furthermore, as we shall see, the dispersion relation of these waves at high
wavenumbers can also be easily manipulated, allowing easy investigations of 
the effects of such changes on horizon effects. 

\section{The model}\label{model}

We shall begin with the simplest form of the model, in which we assume a 
shallow liquid over a flat, horizontal bottom. 
Furthermore, the forces on the liquid will be assumed to be such that they 
allow for a purely horizontal stationary flow profile resulting in a constant 
height (i.e.\  horizontal surface) of the liquid.
Later we shall relax both of these assumptions. 
In addition, we shall assume that the liquid is viscosity free, 
incompressible and irrotational in its flow.

\begin{figure}[ht]
\centerline{\mbox{\epsfxsize=8.5cm\epsffile{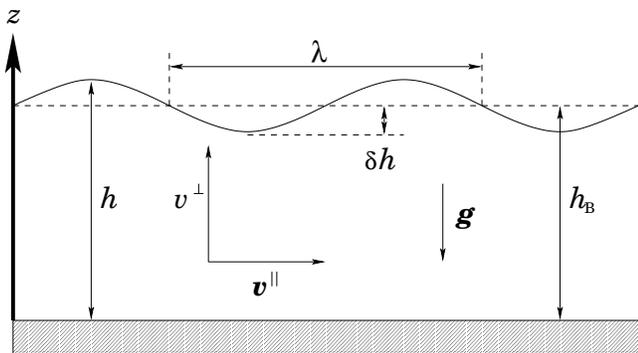}}}
\caption
{
Picture of a gravity wave in the basin and the relevant parameters.
The assumed relation of the dimensions for {\em long} gravity waves
$\delta h \ll h_{\rm B} \ll \lambda$ is not reproduced for the sake of 
conciseness.
}
\label{wave}
\end{figure}

In such a case the density of the liquid remains constant 
($\varrho=\rm const$) and in terms of its local velocity $\vau$
the equation of continuity assumes the simple form
\bea
\label{cont}
\na\cdot\vau=0
\,.
\ea
If we neglect the viscosity of the fluid its dynamics are governed 
by the non-linear Euler equations, see e.g.\ \cite{landau}
\bea
\label{euler}
\frac{d\vau}{dt}=\dot{\vau}
+\left(\vau\cdot\na\right)\vau=
-\frac{\na{p}}{\varrho}+\f{g}+\frac{\f{f}}{\varrho}
\,,
\ea
with $p$ denoting the pressure and $\f{g}=-g\,\f{e}_z$ the gravitational 
acceleration; and $\f{f}=-\varrho\na_\|V^\|$ is some horizontal force 
necessary to establish the stationary horizontal flow; cf.\ Fig.\ \ref{wave}.

For an irrotational flow profile $\na\times\vau=0$ we may simplify the 
Euler equation (\ref{euler}) via $(\vau\cdot\na)\vau=\na(\vau^2)/2$ 
and introduce a velocity potential $\vau=\na\Phi$ arriving at the
Bernoulli equation
\bea
\label{bernoulli}
\dot\Phi+\frac12\left(\na\Phi\right)^2=-\frac{p}{\varrho}-gz-V^\|
\,.
\ea

The boundary conditions are that the vertical velocity must be zero at the
bottom of the tank, the pressure zero at the displaced surface, and change in
the height of the fluid determined by the vertical velocity 
(cf.\ Fig.\ \ref{wave}) 
\bea 
\label{doth}
v^\perp(z=0)=0
\,,\quad
v^\perp(z=h)=\frac{dh}{dt}=\dot{h}+\left(\vau\cdot\na\right)h
\,,
\eea
and
\bea
\label{press}
p(z=h)=0
\,.
\ea

Now, let us consider perturbations $\delta\vau$ to a background flow 
$\vau_{\rm B}$ 
(which is assumed to be stationary, irrotational, and horizontal)
corresponding to small vertical displacements $\delta h$ of the height 
of the fluid, $h$. 
The background flow will be assumed to obey
\bea
\nabla_\perp\vau_{\rm B}=0
\,,\quad
\vau_{\rm B}=\vau^\|_{\rm B}
\quad\leadsto\quad
\na_\|\cdot\vau_{\rm B}=0\,,
\ea
i.e., $h_{\rm B}=\rm const$, and
\bea
{1\over2}\vau_{\rm B}^2= 
-{p_{\rm B}\over\varrho}-gz-V^\|\,,
\ea
where $p_{\rm B}$ will be given by $g(h_{\rm B}-z)$.

We shall assume that the velocity perturbations are also irrotational, 
so that they are given by a potential, $\delta\Phi$. 
The perturbations of the Bernoulli equation are given by
\bea
\label{linbern}
\dot{\delta\Phi} + \vau^\|_{\rm B}\cdot\na_\|\delta\Phi = 
-{\delta p\over\varrho}
\,.
\eea
The boundary condition (\ref{press}) for the pressure together with
$p_{\rm B}=g(h_{\rm B}-z)$ imply 
\bea
\label{linpress}
\delta p(z=h_{\rm B}) = g\varrho\,\delta h
\,,
\ea
and similarly for the vertical velocity
\bea
\label{lindoth}
\delta v^\perp(z=h_{\rm B})=
\dot{\delta h}+\left(\vau_{\rm B}^\|\cdot\na_\|\right)\delta h
\,,
\ea
as well as $\delta v^\perp(z=0)=0$.

It is useful to expand the perturbation potential $\delta\Phi$ 
into a Taylor series
\bea
\label{taylor}
\delta\Phi(x,y,z) = \sum\limits_{n=0}^{\infty}
\frac{z^n}{n!}\delta\Phi_{(n)}(x,y)
\,.
\ea
The boundary condition in Eq.\ (\ref{doth}) implies $\delta\Phi_{(1)}=0$.
Another constraint arises from the equation of continuity (\ref{cont})
\bea
\label{constraint}
\na_\|^2\delta\Phi_{(0)}+\delta\Phi_{(2)}=0
\,,
\ea
and so on for larger values of $n$.
We assume that the wavelength $\lambda$ of the perturbation is much longer 
than the depth $h_{\rm B}$.
In this case the higher-order terms in the Taylor expansion (\ref{taylor}) 
are suppressed by powers of $h_{\rm B}/\lambda\ll1$ since we have 
$\na_\|^2={\cal O}(1/\lambda^2)$.
Keeping only the two lowest (non-trivial) terms in Eq.\ (\ref{constraint})
we find that 
\bea
\label{zett}
\delta v^\perp(z=h_{\rm B})=-h_{\rm B}\na_\|^2\delta\Phi_{(0)}
\,.
\ea
This enables us to combine Eqs.\ (\ref{linbern}), (\ref{linpress}), and
(\ref{lindoth})  
\bea
\label{KFG}
\left(\frac{\partial}{\partial t} +\vau_{\rm B}^\|\cdot\na_\|\right)^2
\delta\Phi_{(0)} 
-gh_{\rm B}\na^2_\|\delta\Phi_{(0)} =0
\,.
\eea
This wave equation, however, equals the Klein-Fock-Gordon (KFG) equation 
\bea
\Box\,\delta\Phi_{(0)}
=\frac{1}{\sqrt{-g}}\partial_\mu
\left(
\sqrt{-g}\,g^{\mu\nu}\partial_\nu\,\delta\Phi_{(0)}
\right)
=0\,, 
\ea
with the effective metric 
(remember $\na_\|\cdot\vau_{\rm B}^\|=0$ and $gh_{\rm B}=\rm const$)
\bea
\label{metric}
{\mathfrak g}^{\mu\nu}_{\rm eff}
=
\left(
\begin{array}{ccc}
1 & \; & \vau_{\rm B}^\| \\ 
\vau_{\rm B}^\| & \; & \vau_{\rm B}^\|\otimes\vau_{\rm B}^\|-gh_{\rm B}\f{1} 
\end{array}
\right)
\,.
\ea
Except for replacing the velocity of the gravity waves $\sqrt{gh_{\rm B}}$ 
by the speed of sound it is exactly the same effective metric as for the 
sonic analogs. 
Calculating the inverse ${\mathfrak g}_{\mu\nu}^{\rm eff}$ 
of the effective metric one obtains
\bea
{\mathfrak g}_{00}^{\rm eff}=
1-\left(\frac{\vau_{\rm B}^\|}{\sqrt{gh_{\rm B}}}\right)^2
\,.
\ea
As one would expect, the condition of an ergosphere 
${\mathfrak g}_{00}^{\rm eff}=0$
is fulfilled where the velocity of the fluid $\vau_{\rm B}^\|$ 
equals the speed of the gravity waves $\sqrt{gh_{\rm B}}$.

\section{Arbitrary Bottom and Height}\label{arbitrary}

Let us now relax the previous assumptions that the bottom and the background
flow surface are both flat and parallel. We introduce arbitrary coordinates on
the bottom and define a vertical coordinate $z$ as orthogonal to the bottom of
the container and geodesic. The spatial metric can always be cast into  the 
form
\bea
d\f{r}^2 = dz^2 + \eta_{ij} dx^i dx^j
\,,
\eea
where $ij$ go over values 1 and 2 (Einstein summation convention), 
and represent the coordinates within the bottom of the container. 
The equation of continuity (\ref{cont}) is now
\bea
\partial_i\left(\sqrt{\eta}\,v^i\right) 
+\partial_z\left(\sqrt{\eta}\,v^z\right)=0
\,,
\eea
with $\eta=\det(\eta_{ij})$ and, assuming irrotational flow, 
the Bernoulli equation (\ref{bernoulli}) becomes
\bea
\dot\Phi + \eta^{ij} (\partial_i\Phi) (\partial_j\Phi) 
+(\partial_z\Phi)^2
=-{p\over\varrho}-V(x^i,z)
\,,
\eea
with
\bea
v_i=\eta_{ij}v^j=\partial_i\Phi
\,,
\ea
and
\bea
v_z=v^z=\partial_z\Phi
\,.
\eea
Here the potential $V(x^i,z)$ already includes the gravitational 
acceleration -- in contrast to the potential $V^\|$ used in the previous 
Section.

The surface of the liquid, defined by $z=h(x^i)$, is where the pressure 
goes to zero and obeys, cf.\ Eq.\ (\ref{doth})
\bea
\label{newdoth}
\dot h + v^i\partial_i h =v^z
\,,
\eea
where $v^i$ and $v^z$ are evaluated at the surface.

Let us now expand these expressions in powers of the vertical height $z$ 
above the bottom at $z=0$. 
The velocity potential $\Phi$, the metric $\eta^{ij}$, 
and the potential $V$ can be written as
\bea
\Phi(x^i,z) 
&=& 
\Phi_{(0)}(x^i) + {z^2\over 2} \Phi_{(2)}(x^i) + {\cal O}(z^3)\,,
\nn
\eta^{ij}(x^i,z) 
&=& 
\eta^{ij}_{(0)}(x^i)+z\,\eta^{ij}_{(1)}(x^i) + {\cal O}(z^2)\,,
\ea
and
\bea
V(x^i,z) = V_{(0)}(x^i) + z g_z(x^i) + {\cal O}(z^2)
\,,
\eea
where we already have incorporated the boundary condition $v_z(z=0)=0$
and introduced the gravitational acceleration perpendicular to the bottom
$g_z=V_{(1)}$.
Similarly we obtain for the pressure $p$
\bea
p = (h-z)p_{(1)} + {\cal O}([h-z]^2)
\,.
\eea
In analogy to the previous Section we assume the height $h$ of the 
fluid to be much smaller than the horizontal length scales on which
the features of the flow profile (e.g.\ $\Phi$, $\eta^{ij}$, $V$, and $p$) 
change significantly -- such as the wavelength $\lambda$.  
In this long-wavelength limit the higher-order terms of the above Taylor
expansions are supressed by powers of $h/\lambda$ and thus can be neglected.

The continuity equation (\ref{cont}) enforces again
\bea
{1\over\sqrt{\eta_{(0)}}}
\partial_i
\left(
\sqrt{\eta_{(0)}}\,\eta^{ij}_{(0)}\partial_j\Phi_{(0)}
\right) 
+\Phi_{(2)}
&\equiv&
\nn
\na^2_\|\Phi_{(0)} + \Phi_{(2)} 
&=& 
0
\,.
\eea
Evaluated at $z=h$ the equation (\ref{newdoth}) for the height 
in terms of the velocity reads
\bea
\dot h + \eta^{ij}
\left[\partial_i\Phi(x^k,h)\right]
\partial_j h=\partial_z\Phi(x^k,h)
\,,
\eea
which again to lowest order in $z=h$ becomes
\bea
\dot h +\eta^{ij}_{(0)} 
(\partial_i\Phi_{(0)}) \partial_j h
=-h\na^2_\|\Phi_{(0)} +{\cal O}(h^2)
\,,
\eea
or, equivalently,
\bea
\label{conserv}
\dot h + {1\over\sqrt{\eta_{(0)}}} 
\partial_i 
\left(
\sqrt{\eta_{(0)}}\,\eta^{ij}_{(0)}h\,
\partial_j\Phi_{(0)}
\right)
=0
\,,
\eea
which can be interpreted as an effective conservation law 
$\dot h + \na_\|\cdot(h\vau^\|)=0$.

The Bernoulli equation transforms into
\bea
\dot\Phi_{(0)} 
&+&
{1\over 2}
\left(\eta^{ij}_{(0)} + z\,\eta^{ij}_{(1)}\right) 
(\partial_i\Phi_{(0)})(\partial_j\Phi_{(0)}) 
\nn
&=& - {p_{(1)}\over\varrho}(h-z)-V_{(0)} - zg_z +{\cal O}(h^2)
\,.
\eea
From the terms linear in $z$ we may infer
\bea
{p_{(1)}\over\varrho}
= g_z + {1\over 2}\eta_{(1)}^{ij}(\partial_i\Phi_{(0)})(\partial_j\Phi_{(0)})
\,.
\eea
We can define an effective gravitational acceleration as 
\bea
\label{effg}
\widetilde g= g_z + 
{1\over 2}\eta_{(1)}^{ij}(\partial_i\Phi_{(0)})(\partial_j\Phi_{(0)})
\,.
\eea
Note that the eigenvalues $\kappa$ of $\eta^{ij}_{(1)}$ with respect to 
$\eta^{ij}_{(0)}$ 
\bea
\eta^{ij}_{(1)} x_j = \kappa\,\eta^{ij}_{(0)} x_j
\,,
\ea
are just twice the inverse of the principle radii of curvature of the 
surface over which the fluid is flowing. 
Thus the extra terms in Eq.\ (\ref{effg}) just represent the vertical 
centrifugal forces on the fluid travel over this curved surface, 
and in all of our following investigations will be negligible.

Let us assume again that we have a stationary background flow which obeys 
these equations, and we are interested in perturbations around this flow. 
The perturbation equations then are
(using $v^i_{\rm B}=\eta_{(0)}^{ij}\partial_j\Phi_{(0)}^{\rm B}$ at $z=0$)
\bea
\dot{\delta h} + {1\over\sqrt{\eta_{(0)}}} 
\partial_i 
\left(
\sqrt{\eta_{(0)}}\,
v^i_{\rm B}
\delta h
\right)
=
\nn
-{1\over\sqrt{\eta_{(0)}}} 
\partial_i 
\left(
\sqrt{\eta_{(0)}}\,\eta^{ij}_{(0)} h_{\rm B}
\partial_j\delta\Phi_{(0)}
\right)
\,,
\eea
and 
\bea
\dot{\delta\Phi}_{(0)} + v^i_{\rm B}\partial_i\delta\Phi_{(0)}
= -\widetilde g\,\delta h
\,.
\eea
We can combine these equations to get
\bea
\left(
\partial_t{1\over\widetilde g}\partial_t
+ 
\partial_i {v_{\rm B}^i\over\widetilde g}\partial_t 
+
\partial_t {v_{\rm B}^i\over\widetilde g}\partial_i 
+
{1\over\sqrt{\eta_{(0)}}}\partial_i
{v_{\rm B}^iv_{\rm B}^j\over\widetilde g}
\sqrt{\eta_{(0)}}\,\partial_j
\right.
\nn
\left.
-
{1\over\sqrt{\eta_{(0)}}}
\partial_i
h_{\rm B}\sqrt{\eta_{(0)}}\,\eta^{ij}_{(0)}
\partial_j
\right)
\delta\Phi_{(0)}
=0\,.
\eea
This again is a KFG equation with a metric given by
\bea
{\mathfrak g}^{\mu\nu}_{\rm eff}
= {1\over h^2_{\rm B}}
\left(
\begin{array}{ccc}
{1} & \; & {v^i_{\rm B}} \\ 
{v^j_{\rm B}} & \; & 
{{v^i_{\rm B}}{v^j_{\rm B}}}-\widetilde g h_{\rm B}\eta^{ij}_{(0)}
\end{array}
\right)
\,,
\eea
where both $\widetilde g$ and $h_{\rm B}$ can depend on the coordinates $x^k$.
We can thus sculpt the effective metric within which these waves flow both 
by altering the velocity of the background flow, by changing $\widetilde g$
 from place to place (primarily by sloping the bottom of the tank), 
or by altering the height of the background flow. 
(Of course the background $h_{\rm B}$ is determined by the background flow, 
$\widetilde g$ and the potential $V$.) 
If one has only the gravitational field as a force on the fluid, the slope 
of the bottom can be used to generate a potential $V$, and also, with more 
severe slopes, to change the value of $\widetilde g$ from place to place.

The ergoregion is defined as the zone where, in order to be travelling at
less than the velocity of the wave in the rest frame of the fluid, 
one cannot be standing still in the lab frame. 
This is the region where the velocity of the fluid is higher than the local 
velocity of the wave, $\sqrt{\widetilde g h_{\rm B}}$. 

If we assume that the bottom is flat (so that $\eta^{ij}=\delta^{ij}$), 
and that the flow is driven by changes in $h_{\rm B}$, we have
\bea
{1\over 2} v^2_{\rm B} + g h_{\rm B} = \rm const
\,.
\eea
But $\sqrt{g h_{\rm B}}$ is the speed of the gravity waves. 
Far from the ergosphere, the velocity $v_{\rm B}$ is small, 
so the we have ${\rm const}=gh_{\infty}$. 
Thus at the erogosphere, where the velocity of the fluid is the velocity 
of the waves, we have 
\bea
h={2\over 3} h_{\infty}
\,.
\eea
Similarly, if we assume that we have a sloping bottom designed so that the 
fluid maintains a constant height $h_{\rm B}$ throughout, we would again 
obtain that the ergosphere should be at a point such that 
$V(x^i)-V_{\infty} = \widetilde g h_{\rm B} /2$. 
If this potential arises purely from the gravitational potential due to the 
slope of the bottom, we must have that the bottom would have to be at a 
height of $h/2$ lower than at infinity. 
I.e., it does not take much of slope to the bottom to create the conditions 
necessary for an analog black hole ergosphere to form.

We note that in the above, we have only kept terms to lowest order in 
$z$ or $h$. 
The validity of this approximation is essentially that all horizontal
derivatives have scales which are much larger than $h$, the height of the 
fluid.
Furthermore, the condition that we need only retain the lowest order in the 
metric $\eta^{ij}$ is also that the curvature of the bottom of the tank be 
on scales which are long with respect to $h$. 
I.e., we are in a "shallow water", long wavelength approximation in these
derivations.

\section{Irrotational background flow}\label{irrot}

We can use the equations of motion of the fluid to derive the most general 
rotationally symmetric and locally irrotational backgound flow profile.
Let us assume that the bottom of the tank is defined by the relation
\bea
Z=f(R)\,,
\eea
in the usual cylindrical coordinates $(Z,R,\varphi)$, where $f$ denotes 
some moderately curved function.
Switching to the adapted coordinates $(z,r,\varphi)$ described in the
previous Section ($Z=f(R) \leftrightarrow z=0\,;\,r=R$)
gives the spatial metric
\bea
d\f{r}^2 = dz^2
&+&
dr^2\left(1+f'(r)^2\right)
\left(1-z{f''(r)\over\sqrt{(1+f'(r)^2)^3}}\right)^2 
\nn
&+& 
d\varphi^2
\left(r-z{f'(r)\over\sqrt{1+f'(r)^2}}\right)^2
\,,
\eea
for which the lowest order metric is
\bea
d\f{r}^2_{(0)}=
dz^2+\left(1+f'(r)^2\right)dr^2+r^2 d\varphi^2
\,.
\eea

To lowest order in $h$ the flow equations are the effective continuity
equation (\ref{conserv})
\bea
\partial_r\left(\sqrt{1+f'(r)^2}\,r h\,v^r\right) &=& 0\,,
\ea
the condition for a locally irrotational flow 
\bea
v^\varphi &=& {L\over r^2}\,,
\ea
with $L$ being some constant related to the angular momentum, 
and finally the Bernoulli equation
\bea
\left(1+f'(r)^2\right)(v^r)^2 + r^2 (v^\varphi)^2 &=&
\nn 
-g\frac{h-h_{\infty}}{\sqrt{1+f'(r)^2}} - gf(r)
\,,
\eea
where we have neglected the "centrifugal" term in Eq.\ (\ref{effg}) 
as it will be very small for our situation.
These give
\bea
v^r = {C h_{\infty}\over r h \sqrt{1+f'(r)^2}}
\,,
\ea
and
\bea
{1\over 2}\left({C^2h_{\infty}^2\over h^2r^2} +{L^2\over r^2}\right)
= -g\frac{h-h_{\infty}}{\sqrt{1+f'(r)^2}} -gf(r)
\,.
\eea
Thus, we either need $h$ to change as a function of $r$, or we need a 
non-trivial $f(r)$. 
Choosing $f(r)= -F/r^2$ allows a consistant solution with contant height 
$h=h_{\infty}$ for the fluid as long as $F$ is given by $F=(C^2+L^2)/g$.
The effective metric for the fluid is then of the form
\bea 
\label{metricvariable}
ds_{\rm eff}^2
&=&
\frac{h_{\infty}}{\widetilde g}
\left(\widetilde gh_{\infty}-\frac{C^2+L^2}{r^2}\right)dt^2
\nn
&&
+2\frac{h_{\infty}}{\widetilde g}
\left(
\sqrt{1+f'(r)^2}\,\frac{C}{r}\,dt\,dr
+L\,dt\,d\varphi
\right) 
\nn
&&
-\frac{h_{\infty}}{\widetilde g}
\left(
\left[1+f'(r)^2\right]dr^2
+r^2d\varphi^2
\right)
\,,
\eea
with ${\widetilde g}=g_z=g/\sqrt{1+f'(r)^2}$ denoting the effective 
gravitational acceleration.

In summary the analogy to a curved space-time and the concept of an effective
metric can still be applied in the case of non-horizontal flow provided that
the local variation of the height of the fluid and the slope of its bottom 
are sufficiently small.
Nevertheless, the global changes may well be significant.

A variation of $h$ and $g_z$, the component of the force perpendicular to 
the bottom, does in general also entail a change of the local velocity 
$\sqrt{g_z h}$ of the gravity waves.
Such a spatial dependence may lead to further interesting effects:
In analogy to optics one may introduce an effective index of refraction 
which then also acquires a non-negligible gradient.
In such a situation the gravity waves may be scattered by this gradient 
or even the phenomenon of total reflection could occur.
As we shall see later in Section \ref{kerr}, this mechanism may be one 
ingredient for generating an instability.

\section{Surface tension}\label{surface}

So far we have considered ideal fluids without any internal forces.
However, if we take the surface tension of the liquid into account, 
the pressure at its surface no longer  vanishes. 
Accordingly, the upper boundary condition (\ref{press}) for the pressure 
is modified to
\bea
p(z=h)=-\alpha\na_\|^2h
\,,
\ea
where $\alpha$ denotes the fluid's surface-tension coefficient and 
$\na_\|^2h$ is the curvature of its surface in the linear approximation.
Consequently we obtain 
\bea
\delta p(h_{\rm B})=\varrho g\,\delta h - \alpha\na_\|^2\delta h
\,,
\ea
instead of Eq.\ (\ref{linpress}).
This results in a extra term in the velocity perturbation equation 
(\ref{linbern})
\bea
\dot{\delta\Phi} + \vau^\|_{\rm B}\cdot\na_\|\delta\Phi = 
- g\,\delta h + \frac{\alpha}{\varrho}\na_\|^2\delta h
\,.
\ea
As we shall see below, the effects of surface tension become relevant for 
small wavelengths only.
In this limit we may neglect the variation of the background flow
$\na_\|\otimes\vau_{\rm B}^\|\approx0$ and obtain a modified wave equation
\bea
\left(\frac{\partial}{\partial t} + \vau_{\rm B}^\|\cdot\na_\|\right)^2
\delta\Phi
=
gh_{\rm B}\na_\|^2\delta\Phi
-\frac{\alpha h_{\rm B}}{\varrho}\na_\|^4\delta\Phi
\,.
\ea
In terms of the capillary constant $a^2=\alpha/(\varrho g)$ and the 
velocity of the unperturbed gravity waves $c_{\rm B}^2=gh_{\rm B}$
this wave equation results in the following dispersion relation
\bea
\label{dispersion}
\left(\omega+\vau_{\rm B}^\|\cdot\f{k}\right)^2
=
c_{\rm B}^2\left(\f{k}^2+a^2\f{k}^4\right)
\,.
\ea
Therefore the incorporation of the effects of surface tension leads to a 
``superluminal'' dispersion relation (in the terminology of \cite{laser})
since (for $\vau_{\rm B}^\|=0$) the 
group velocity $d\omega/dk$ as well as the phase velocity $\omega/k$ 
exceed $c_{\rm B}$ for large wavenumbers $k$.

However, we should bear in mind that the above calculations are still based 
on the assumption of {\em long} gravity waves $\lambda \gg h_{\rm B}$.
For short gravity waves $\lambda \ll h_{\rm B}$, on the other hand, 
the dispersion relation reads (for $\vau_{\rm B}^\|=0$, cf.\ \cite{landau})
\bea
\omega^2=gk(1+a^2k^2)
\,.
\ea
Hence we can use the  ratio $a/h_{\rm B}$ in order to alter the 
dispersion relation for large wavenumbers $k$: 
For $a \gg h_{\rm B}$ the capillary waves dominate before the wavelength 
becomes smaller than the height, and we have  a superluminal dispersion 
relation, whereas for $a \ll h_{\rm B}$ the short gravity waves dominate 
before the surface tension becomes important, 
and thus one initially has a subluminal dispersion relation, before the
capillary waves finally take over at very short wavelengths.

For example,  for mercury the surface tension coefficient $\alpha$ is about 
$\alpha\approx0.46\;{\rm N}\;{\rm m}^{-1}$ at room temperature
$293\;\rm K$ and hence its capillary constant $a\approx1.9\;\rm mm$.
For water at $293\;\rm K$ we have
$\alpha\approx0.0725\;{\rm N}\;{\rm m}^{-1}$ and hence
$a\approx2.7\;\rm mm$.
This quantity can easily be manipulated by changing the temperature, 
adding surfactants, or by changing the fluid used.

\section{Viscosity}\label{viscosity}

The dynamics of a viscous but still incompressible fluid are governed 
by the Navier-Stokes equations 
\bea
\label{navier}
\frac{d\vau}{dt}=\dot{\vau}
+\left(\vau\cdot\na\right)\vau=-\frac{\na{p}}{\varrho}+\f{g}
+\nu
\na^2\vau
\,,
\ea
where $\varrho\nu$ denotes the dynamic viscosity of the liquid 
and $\nu$ its kinematic viscosity.

The boundary conditions have to be modified as well.
Instead of Eq.\ (\ref{press}) we have now
\bea
\label{vis-press}
p(z=h)=2\varrho\nu\partial_zv_z(z=h)
\,,
\ea
and there are two additional restrictions on $\vau_\|$
\bea
\label{vis-vel}
\partial_z\vau_\|(z=h) &=& - \na_\| v_z(z=h)
\,,
\ea
and
\bea
\label{stick}
\vau_\|(z=0) &=& 0
\,.
\ea

Let us investigate the effects of a finite but small viscosity on the 
wave propagation -- where we restrict our examination to the case of
a vanishing background flow $\vau_{\rm B}=0$ for simplicity and employ 
the plane-wave ansatz with a frequency $\omega$ and a wavenumber $\f{k}$.

For an incompressible fluid the divergence of the linearized 
Eq.\ (\ref{navier}) yields (for $\vau_{\rm B}=0$)
\bea
\label{laplace}
\na^2 \delta p = \left( \partial_z^2 - \f{k}^2 \right) \delta p = 0
\,,
\ea
which has the solution $\delta p=A\cosh(kz)+B\sinh(kz)$ with $k=|\f{k}|$.

In the long-wavelength limit $kh_{\rm B}\ll1$ we may approximate 
$\sinh(kz) \approx kz$ and
$\cosh(kz) \approx 1+(kz)^2/2$.
Inserting the resulting expression back into the Navier-Stokes equations 
(\ref{navier}) yields 
\bea
\left( \partial_z^2 - k^2 - i\frac{\omega}{\nu} \right) \delta v_z =
\frac{Ak^2z+Bk}{\varrho\nu}
\,.
\ea
Defining $\widetilde{k}$ via $\widetilde{k}^2=k^2+i\omega/\nu$ and
$\Re(\widetilde{k})>0$ the general solution of this equation can be 
written as
\bea
\label{vis-sol}
\delta v_z = -\frac{Ak^2z+Bk}{\varrho\nu\widetilde{k}^2}
+Ce^{+\widetilde{k}z}+De^{-\widetilde{k}z}
\,.
\ea

In addition to the long-wavelength limit $\lambda \gg h_{\rm B}$ 
we assume the viscosity $\nu$ to be very small $\nu \ll \omega h_{\rm B}^2$.
In this case the exponentials $\exp(\pm\widetilde{k}z)$ are very rapidly 
varying functions -- which can be used to simplify the analysis.

Combining the equation of continuity 
$\partial_z\delta v_z+i\f{k}\cdot\delta\vau_\|=0$
with the velocity boundary condition in Eq.\ (\ref{vis-vel}) one obtains
$(\partial_z^2+k^2)\delta v_z(z=h_{\rm B})=0$.
In view of $|\widetilde{k}| \gg k$ and $\widetilde{k}h_{\rm B} \gg 1$ 
this implies that $C$ is extremely small
$C\propto\exp(-\widetilde{k}h_{\rm B})/\widetilde{k}^2$ 
and thus can be neglected.

On the other hand, from $\delta v_z(z=0)=0$ we obtain 
$D=Bk/(\varrho\nu\widetilde{k}^2)$.
Therefore, the term $D\exp(-\widetilde{k}z)$ in Eq.\ (\ref{vis-sol})
is relevant in a very thin boundary layer of order $\sqrt{\nu/\omega}$ 
over the bottom only, cf.\ \cite{landau}. 

In analogy to Eq.\ (\ref{linpress}) we may linearize the boundary condition 
for the pressure in Eq.\ (\ref{vis-press}) which fixes the integration 
constant $A$.

The remaining condition $\vau_\|(z=0)=0$, i.e.\ $\partial_z\delta v_z(z=0)=0$,
can be used to eliminate $D$ and hence $B$.
As one might expect, the solution for $\delta v_z$ in the presence of a small
viscosity displays only slight deviations 
($B$ is of order $\sqrt{\nu}$) from the linear profile 
$\delta v_z \propto z$ used in the previous Sections --
as long as one is well above the aforementioned boundary layer. 

Finally, Eq.\ (\ref{doth}), i.e.\ $\delta v_z(z=h_{\rm B})=i\omega\delta h$, 
enables us to derive the dispersion relation
\bea
\label{vis-disp}
\omega^2=gh_{\rm B}k^2-gk^2\sqrt{\frac{\nu}{i\,\omega}}+{\cal O}(\nu)
\,.
\ea
Here one can read off the characteristic damping time $\tau$
after which the viscosity effects become significant
\bea
\label{tau-long}
\tau\sim\frac{h_{\rm B}}{\sqrt{\nu\omega}}
\,.
\ea
One observes that high frequencies are damped faster.
This tendency becomes much stronger in the r\'egime 
of short gravity waves where  
\bea
\label{tau-short}
\tau\sim\frac{1}{\nu k^2}\sim\frac{g^2}{\nu\omega^4}
\ea
holds, see e.g.\ \cite{landau}.
As a result, the at a first glance undesirable effects of viscosity
can be utilized to damp out potential high-frequency noise and so
single out the interesting (medium-wavenumber) instabilities
by tuning $\nu$.

For example, water at room temperature has a kinematic viscosity of
$\nu\approx1\;{\rm mm}^{2}\,{\rm s}^{-1}$.
Assuming a height $h_{\rm B}=10\;{\rm cm}$ and a frequency 
$\omega=1\;{\rm Hz}$ 
we infer from Eq.\ (\ref{tau-long}) a damping time of the order of 
one minute -- which still seems to be sufficient.

The kinematic effect of the friction within the liquid could be diminished 
by selecting a fluid with a relatively high density $\varrho$
(e.g.\ mercury\footnote{
The use of a fluid like mercury has just another 
advantage since one would be able to detect the gravity waves very easily 
via considering the reflection at the fluid's surface in this case.}) 
in comparison with its inner viscosity.
For mercury at room temperature $\nu$ is given by 
$\nu\approx0.12\;{\rm mm}^{2}\,{\rm s}^{-1}$. 

However, there is also another problem induced by the finite viscosity:
the ansatz for the $z$-independent background flow profile used in the 
previous Sections is not appropriate anymore since the fluid sticks to
the bottom of the basin, cf.\ Eq.\ (\ref{stick}).

One way to solve this problem is to move the bottom so that its velocity 
is the same as that of the fluid -- at least in the interesting region, 
for example near the horizon. 
(One might also imagine manipulating the fluid near the bottom and thereby
effectively simulating this motion.)
After transforming into its rest frame the above calculation demonstrates 
that the solutions derived in the previous Sections are still a very good 
approximation.

Without moving the bottom the boundary condition in Eq.\ (\ref{stick}) 
enforces a significant $z$-dependence of the background flow profile.
For instance a constant horizontal force 
(needed for maintaining the stationary flow) 
implies a parabolic flow profile $v^z_{\rm B} \propto z(2h_{\rm B}-z)$.
Unfortunately, it is not possible to cast the full wave-equation into an as 
tractable form as in Eq.\ (\ref{KFG}) allowing for the identification of an 
effective metric in this situation 
(this problem is currently under investigation).
The main obstacle is that the flow is longer irrotational. 

Nevertheless, for other scenarios one might be able to overcome this 
difficulty. 
If one injects a nearly $z$-independent stationary inflow 
(driven by a turbine, for example) on one side of the basin,
then the flow will basically remain irrotational throughout the basin
-- outside a thin Prandtl boundary layer at the bottom, 
cf.\ \cite{landau}.

In view of the relatively large velocities involved and the supposedly
small viscosity this layer may well be turbulent.
However, by an appropriate preparation of the bottom's surface 
(e.g.\ dolphin skin effect) the indued drag can be diminished.

Since the properties of the flow outside the thin boundary layer are nearly 
the same as in the case without viscosity and in view of the remarks after
Eq.\ (\ref{vis-sol}) one would expect that the basic properties of the gravity
waves as discussed in the previous Sections are not drastically affected 
by a small internal friction in this case.

Beside the scenarios described above there is also another -- more exotic -- 
solution for the viscosity problem conceivable:
a superfluid does not stick to the boundary and its vorticity is
quantized. But the necessity of using gravity waves in say liquid Helium 
makes the experimental realization far more difficult than is desirable.

\section{Energy}\label{energy}

There are two different concepts of a metric in our model:
Firstly, the Minkowski metric determining the length- and time-scales
within our laboratory; 
and, secondly, the effective metric -- which is experienced by the
gravity waves only.

These two concepts lead to two distinct notions of energy:
For time-independent external forces (and inviscid fluids) the Noether 
theorem demands the conservation of the total energy of the flow.
In addition, assuming a stationary background flow profile, 
we may find a conserved energy associated to the gravity waves:

At the boundary $\partial G$ of the basin, the normal component of the 
velocity $\vau\cdot\f{n}$ has to vanish. 
In terms of the velocity potential $\Phi$ this corresponds to  
Neumann boundary conditions 
\bea
\vau\cdot\f{n}\mid_{\partial G}=0
\quad\leadsto\quad
\f{n}\cdot\na\Phi\mid_{\partial G}=0
\,.
\ea
This enables us to accomplish a spatial integration by parts and in 
complete analogy to the 2+1 dimensional curved space-time one may derive
a conserved energy 
\bea
E=\int d\Sigma_\mu\,T^{\mu\nu}\,\xi_\nu=\int d^2r\,T^0_0
\,,
\ea
where $\xi^\mu=\partial/\partial t$ denotes the Killing vector associated to 
the physical laboratory time and $\Sigma_\mu$ the spatial hyper-surface. 
For the effective metric in Eq.\ (\ref{metric}) we obtain the energy density
\bea
\label{T00}
T^0_0=\frac{1}{2}\left[
\left(\frac{\partial\phi}{\partial t}\right)^2+
gh_{\rm B}\left(\na_\|\phi\right)^2
-\left(\vau_{\rm B}^\|\cdot\na_\|\phi\right)^2
\right]
\ea
of the perturbations $\phi=\delta\Phi_{(0)}$, 
i.e.\ gravity waves (see also \cite{stone}).

We observe that this energy density contains negative parts inside the 
ergoregion $|\vau_{\rm B}^\||>\sqrt{gh_{\rm B}}$. 
This observation points to the possibility of an instability associated to
the ergoregion.
(A positive definite conserved energy density would prove stability.)
The total energy of the fluid including the background flow is of course 
always positive.

Note that the energy conservation law derived above is violated if $G$ 
has a hole, such as at the drain, etc.
This problem, however, arises for real black holes as well.  

\section{Non-rotating black hole}\label{black}

If we neglect the small slope $f'(r)\ll1$ of the bottom 
(cf.\ the remarks at the end of Section \ref{arbitrary})
in Eq.\ (\ref{metricvariable}) the constant quantities $g$ and $h_{\rm B}$
can be absorbed by a simple rescaling and we arrive at
\bea
\label{prekerr} 
ds_{\rm eff}^2
&=&
\left(gh_{\infty}-\frac{C^2+L^2}{r^2}\right)dt^2
+2\frac{C}{r}\,dt\,dr
+2L\,dt\,d\varphi
\nn
&&
-dr^2-r^2d\varphi^2
\,.
\eea
If we take $L=0$, we exactly recover a
Painlev{\'e}-Gullstrand-Lema{\^\i}tre (PGL) type metric \cite{PGL}
\bea
\label{PGL}
ds^2_{\rm eff}
=\left[c_{\rm B}^2-w^2(r)\right]dt^2+2w(r)\,dt\,dr-dr^2-r^2d\varphi^2
\,,
\ea
with $c_{\rm B}^2= gh_{\infty}$ and $w(r) =\sqrt{1+f'(r)^2}\,v^r=C/r$.
As it is well-known, by means of the singular coordinate transformation 
\bea
dt \to d\tilde t=dt+dr\,\frac{w(r)}{c_{\rm B}^2-w^2(r)}
\,,
\ea
the stationary PGL metric can be cast into the static Schwarzschild form 
\bea
ds^2_{\rm eff}=
\left(c_{\rm B}^2-w^2\right)d{\tilde t\,}^2-
\frac{c_{\rm B}^2}{c_{\rm B}^2-w^2}dr^2
-r^2d\varphi^2
\,.
\ea
Obviously the horizon occurs when $w^2=c_{\rm B}^2=gh_{\rm B}$, i.e.\ when the 
velocity of the (radially) flowing fluid exceeds the speed of the (long) 
gravity waves $\sqrt{gh_{\rm B}}$.
An inward flowing liquid $w<0$ simulates a black hole whereas an outward
flow $w>0$ evidently corresponds to a white hole. 
The black hole branch can be used to observe the inversion of the centrifugal 
acceleration \cite{centrifugal} mentioned in the Introduction and, of course, 
the trapping of the waves inside the horizon $w^2=c_{\rm B}^2$.

\section{White hole}\label{white}

The white hole branch $w>0$ of Eq.\ (\ref{PGL}) offers another interesting 
phenomenon:
As demonstrated in Ref.\ \cite{eardley}, all incident waves pile up at the
horizon (since they cannot penetrate) and get arbitrarily blue-shifted there
-- if one neglects the change in dispersion relation, and thus group
velocity, at high wavenumbers.
In our model, however, the blue-shifted waves eventually leave the 
regime of the long gravity waves.
The subsequent behavior depends on the character of the dispersion relation,
in particular whether it is subluminal or superluminal.
In the latter case the incident 
 waves will eventually  penetrate the horizon, once their wavelength has
 become sufficiently short to alter the group velocity to one larger than
 the (low frequency) velocity of the waves. In the former case, the short
 wavelengths will be swept out by the fluid which is now flowing faster at
 the horizon than the group velocity.   
These features associated with white holes may thus be observed 
experimentally for these gravity wave analogs. Furthermore, since the
dispersion relation can be adjusted by varying the depth and the surface
tension of the liquid, one can study the effect over a wide range of 
physical situations.

\section{Rotating black hole}\label{kerr}

Stationary flow profiles containing a component in $\varphi$-direction 
(vortex solutions with non-zero $L$) can be used to model rotating (Kerr) 
black holes. 
With a further rescaling and a redefinition of the constants $C$ and $L$ 
we may absorb the speed of the gravity waves $gh_\infty$ in 
Eq.\ (\ref{prekerr}) completely and then the corresponding effective 
metric assumes the following form
\bea
\label{kerrmetric}
ds^2_{\rm eff}
&=&
\left(1-\frac{C^2+L^2}{r^2}\right)dt^2+2\frac{C}{r}\,dt\,dr+2L\,dt\,d\varphi
\nn
&&-dr^2-r^2d\varphi^2
\,.
\ea
The space-time structure of a Kerr black hole is more complicated than the
Schwarzschild geometry for there is a difference between the static limit
(or the ergosphere, see e.g.\ \cite{misner}) and the horizon:
The static limit ${\mathfrak g}_{00}=0$ denotes the region beyond 
which no particle can remain at rest. 
This, as mentioned above, corresponds to the surface where
the velocity of the fluid equals the velocity of the waves, 
i.e.\ $r^2=C^2+L^2$. 
The horizon is the ``point of no return'', and for an axially symmetric
flow, corresponds to the surface where the radial flow velocity equals
the velocity of the waves $r^2=C^2$.

The region between these two critical points, in both Kerr space-time and 
this model, is called the ergoregion and allows for the occurrence of the 
superradiant modes. 

According to Eq.\ (\ref{T00}) the energy density may become negative inside
the ergoregion $r^2<C^2+L^2$.
As already anticipated in Section \ref{energy}, this observation can be 
interpreted as an indicator of instability. 
Indeed, in complete analogy with the Kerr black hole this analog should 
exhibits the phenomenon of superradiance:
An incident wave with non-vanishing angular momentum scatters from the
region around the black hole (analog) -- i.e.\ the vortex -- and the 
amplitude of the reflected wave is larger than that of the ingoing wave.
The necessary energy is extracted from the rotational energy of the 
background.

Since the metric in Eq.\ (\ref{kerrmetric}) possesses two independent Killing 
vectors, $\partial/\partial t$ and $\partial/\partial\varphi$ we may 
find a complete set of solutions of the wave (KFG) equation 
\bea
\label{full}
\left(
\partial_t^2+
2\frac{C}{r}\,\partial_t\partial_r+
2\frac{L}{r^2}\,\partial_t\partial_\varphi+
\frac{CL}{r^3}\,\partial_r\partial_\varphi+
\frac{1}{r}\partial_r\frac{CL}{r^2}\,\partial_\varphi
\right.
\nn
\left.
+\frac{1}{r}\,\partial_r\left(\frac{C^2}{r^2}-1\right)r\,\partial_r+
\left(\frac{L^2}{r^4}-\frac{1}{r^2}\right)\partial_\varphi^2
\right)\phi=0
\,,
\ea
by the following separation ansatz 
\bea
\phi(t,r,\varphi)=\exp\left\{ -i \omega t + i m \varphi \right\}
\phi_{\omega m}(r)
\,.
\ea
Note that an analogous separation ansatz in $t,\vartheta,\varphi$ 
is even possible in the real 3+1 dimensional Kerr metric -- 
which is a less trivial statement.

The remaining function $\phi_{\omega m}(r)$ obeys a second-order ordinary
differential equation.
In terms of the Regge-Wheeler tortoise coordinate defined by
\bea
dr_*=\frac{r^2}{r^2-C^2}\,dr
\,,
\ea
the KFG equation (\ref{full}) at the horizon $r^2=C^2$, i.e.,
$r_*=-\infty$, simplifies to
\bea
\left(\partial_{r_*}+2i\left[\omega-\frac{L}{C^2}\,m\right]\right)
\partial_{r_*}\phi_{\omega m}=0
\,.
\ea
The two linearly independent solutions to this equation, i.e.,
(approximately) constant and purely oscillating, respectively, 
correspond to ingoing and outgoing modes, respectively.  

In analogy to the Kerr metric we may introduce the angular velocity of 
horizon
\bea
\Omega_{\rm H}=\frac{L}{C^2}
\,.
\ea
For a Kerr black hole this quantity is 
bounded\footnote{Otherwise the metric would describe a naked singularity
without ergoregion and horizon, etc.} 
by its total mass $M$ via $2\Omega_{\rm H}<1/M$.
With the gravity wave analogs, however, it is possible to generate
rather large values of $2\Omega_{\rm H}$ 
-- in particular if one allows for non-negligible slopes -- since 
$C$ and $L$ can be varied independently.

The Wronskian associated to the second-order ordinary differential equation
reads at the horizon
\bea
W_{\omega m}[\phi]=\phi^*\partial_{r_*}\phi-\phi\,\partial_{r_*}\phi^*
+2i(\omega-\Omega_{\rm H}m)\phi^*\phi
\,.
\ea
Inserting the ingoing mode $\partial_{r_*}\phi=0$ we obtain
\bea
W_{\omega m}[\phi]=
2i(\omega-\Omega_{\rm H}m)\left|{\cal T}_{\omega m}\right|^2
\,,
\ea
where $|{\cal T}_{\omega m}|^2=|\phi_{\omega m}|^2$ denotes the 
transmission coefficient. 
Since we have restricted our solution to be purely ingoing at the horizon,
it will contain ingoing as well as outgoing components
(determined by the reflection coefficient ${\cal R}_{\omega m}$) 
at spatial infinity $r=\infty$ in general
$\phi\propto\exp(-i\omega r) + {\cal R}_{\omega m}\exp(i\omega r)$.
In this limit the wave equation (\ref{full}) reduces to the usual form 
and hence the associated Wronskian reads
\bea
W_{\omega m}[\phi]=
2i\omega\left(1-\left|{\cal R}_{\omega m}\right|^2\right)
\,.
\ea
For a regular second-order ordinary differential equation on a contiguous
interval the Wronskian is conserved which implies \cite{super} the following 
relation between the transmission ${\cal T}_{\omega m}$ and 
the reflection ${\cal R}_{\omega m}$ coefficients
\bea
1-\left|{\cal R}_{\omega m}\right|^2=
\frac{\omega-m\Omega_{\rm H}}{\omega}
\left|{\cal T}_{\omega m}\right|^2
\,.
\ea
As one can easily infer from the equation above, if
$m\Omega_{\rm H}>\omega$ the reflection coefficient is greater than one
$|{\cal R}_{\omega m}|>1$, i.e. the scattered wave has a larger amplitude
than the incident one.
This amplification process corresponds to the phenomenon of superradiance.

However, for a massless scalar field in the asymptotically flat Kerr geometry 
the scattered wave escapes to infinity. To make an amplifier unstable,
feedback is required, see \cite{kerrstable}. 
That outgoing wave must be reflected back toward the hole for repeated 
amplification.
This is possible if the field has a finite mass \cite{kerrunstable} 
or if one, for example, 
encloses the rotating black hole by a large spherical mirror \cite{bomb}.

As explained in Section \ref{irrot}, for the gravity wave analogs
there is another mechanism which may force the scattered wave to ``come back''
-- the total internal reflection.
Since the velocity of the wave is $\sqrt{gh}$, if one can manipulate $h$, 
as happens  for example near the drain of a bath tub, the change in the 
effective refractive index may entail effectively trapping the  waves -- 
which then can be amplified via superradiant scattering without bound, 
or, rather, until non-linear effects dominate.  

In summary it is possible that some of the features which one regularly
observes in the vortex flow near the drain of one's bathtub are in fact
the result of instabilities which are exactly analogous to the behaviour
of waves near a (rotating) black hole.
It should be mentioned here that the arguments above do not explain the 
formation of the vortex, but refer to its linear instability.

\section{Inner horizon}\label{inner}

The instability indicated by negative parts of the energy density in Eq.\ 
(\ref{T00}) is not necessarily restricted to rotating black holes:
In the presence of an inner horizon in addition to the outer one,
such as occurring in the Reissner-Nordstr\"om metric, and for an 
altered dispersion relation, runaway solutions can exist even in the 
purely one dimensional flow.

This somewhat surprising fact has been demonstrated in Ref.\ \cite{laser}.
In the following we give a brief repetition of the basic explanation adapted 
to subluminal high frequency dispersion relation. Note that the terms
"sub-" or "superluminal" do not, despite their origin, refer to light,
but rather to whether or not the group velocity at high wavenumbers is
less than or greater than it is at very low wavenumber (long wavelength). 
They will also refer to regions where the velocity of the fluid is smaller 
or larger than than the low-wavenumber velocity of the waves.

We shall restrict ourselves to a one-dimensional flow. We consider a case
where there are two horizons, one a white hole (where the velocity of the
fluid drops below the low wavenumber velocity of the wave) and the other a
black hole (where the velocity of the fluid goes above the low wavenumber
group velocity). 
It turns out (for not entirely well understood 
reasons\footnote{The 1+1 dimensional scalar wave equation without 
dispersion is conformally invariant which explains the absence of 
any mixing in this case. For a non-trivial dispersion, however,
this problem is less clear.})
that although there are two branches of the dispersion curve (for a static
fluid these correspond to the two directions of propagation of the wave),
wave packets remain on one or the other of these branches even when
interacting with horizons, and spatially varying flows. There seems to be
very little backscatter from one branch to the other. We shall restrict
attention to the one branch which at low wavenumbers represents waves
which are travelling in a direction opposite to the fluid flow. 

\begin{figure}
\centerline{\mbox{\epsfxsize=9.5cm\epsffile{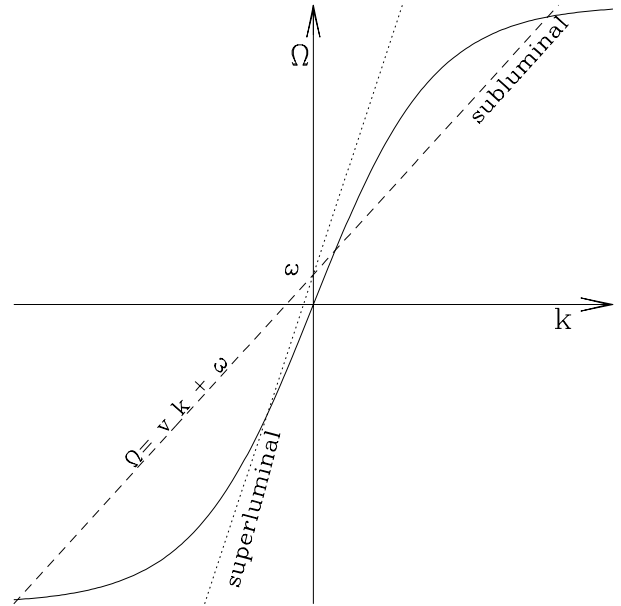}}}
\caption{
One branch of the dispersion relation where the high wavenumber group 
velocity is smaller than the low frequency. 
Plotted are the two lines of $\Omega$ vs.\ $k$ for different velocities, 
but the same value of the lab frame frequency $\omega$.
Note the intersection points of these straight lines with the dispersion 
curve indicating the possible values or $k$ for the given $\omega$ at the 
given fluid velocity. 
}
\label{figdisp}
\end{figure}

Assuming adiabatic motion of this wave packet
(in analogy to the geometric optics approximation), 
the effect of the fluid flow is to alter the dispersion relation to
\bea
\omega +v k = F(k)
\,,
\eea
where the mode has the form $e^{i(\omega t +kx)}$. 
I.e., $\omega$ is the frequency of the wave in the lab frame, 
not the rest frame of the fluid. 
Define $\Omega=\omega+vk$ as the frequency in the fluid rest frame. 
In Fig.\ \ref{figdisp} we sketch the subluminal dispersion relation 
as a plot of $\Omega$ versus $k$. 
To determine the possible values of $k$ for any given value of $\omega$, 
we can plot the line $\Omega-vk=\omega$ and look for the intercepts 
with the dispersion curve.
We shall concentrate on small  values of $\omega$, which is also the 
intercept of the line with the $\Omega$ axis. 

On the subluminal side of the horizon, where $v$ and thus the slope of the 
line is less than the dispersion slope at $k=0$, 
there will in general be three points of intersection, 
and with very different group velocities. 
The one near the origin $k=0$ has a group velocity, 
\bea
v_{\rm g}=-\frac{d\omega}{dk}=v-\frac{dF}{dk}
\,,
\ea
which is negative, 
corresponding to travel to the left, against the flow of the current. 
The other two, one at positive $\Omega$ and one at negative $\Omega$ 
both have positive group velocity and thus correspond to packet travel 
with the current flow.
(Their rest frame group velocity is so small that even the subluminal 
current is sufficiently fast to drag them along). 

Because the background current flow is assumed to be stationary, the
frequency $\omega$ is conserved during the motion of a wave packet.
Let us consider what the possible wavenumbers are for this same value of 
$\omega$ on the superluminal side. 
The slope of the line is now much greater than the low wavenumber slope of the
dispersion curve, and there is now only one intersection with the dispersion
curve, at negative values of $\Omega$. Furthermore the group velocity 
here is positive.
These wave packets, even though they have low wavenumber, 
are dragged along with the fluid and  
travel in the same direction as the fluid. 

One can, as for scalar waves, define a ``norm'' of the various wave 
packets by the conserved Klein-Fock-Gordon inner product
\bea
\inner{\Phi}{\Phi} 
&=& 
\frac{i}{2}\int d\Sigma^\mu\left(
\Phi^*{\partial}_\mu\Phi-\Phi{\partial}_\mu\Phi^*\right)
\nn
&=&
\frac{i}{2}\int d\Sigma^\mu\,
\Phi^*\stackrel{\leftrightarrow}{\partial}_\mu\Phi
\nn
&=&
\frac{i}{2}\int dx\,\Phi^*
\left(\stackrel{\leftrightarrow}{\partial}_t-v(x) 
\stackrel{\leftrightarrow}{\partial}_x\right)\Phi
\nn
&=&
\Omega(x) \int dx\,|\Phi|^2
\,,
\eea
where we have assumed the wave packet to be sufficiently localized that 
$v$ is constant over the packet. 
I.e., the sign of the norm is the same as the
sign of $\Omega$. Thus, for positive $\omega$ the wave packets on the
superluminal side will have negative norm, while those on the subluminal
side can be either positive or negative norm (two are
positive and one is negative).

Let us consider a wave paket with positive norm, low value of $\omega$ but
large value of $k$ being dragged toward the black hole horizon. 
As it comes closer, $v$ increases, and the wavelength of the packet is 
stretched out, with $k$ becoming smaller and smaller. 
As it hits the horizon, where the slope of the line becomes essentially 
the same as that of the low wavenumber dispersion curve, 
the adiabatic assumption (geometric optics approximation) fails, and one gets 
a mixing of the various possible values of $k$ for the given $\omega$.
However, whatever happens, the wave packet must thereafter leave the
horizon, either on the subluminal side, or the superluminal. On the
subluminal side the only possibility is that the wave  exit with the
small value of $k$ near zero, as this is the only one of the three
possibilities which travels away from the horizon on the subluminal side.
On the superluminal side, the only possibility is that the wave packet
leave with negative norm. 
The fluid flow at the horizon will have mixed positive with
negative norm solutions. 
This is precisely the requirement in the quantum system that particle 
production take place, since this mixing of positive and negative norms 
is precisely what leads to non-trivial Bogoliubov coefficients
\bea
\beta=\inner{\Phi^*_{\rm in}}{\Phi_{\rm out}}
\,.
\ea
For our purpose, what this means is that the wave packet
which leaves the horizon on the subluminal side must have a larger norm
than did the wave packet which was dragged toward the horizon. (Since the
total norm is conserved, and since the norm of the packet on the
superluminal side is negative, the subluminal packet must be larger.)

\begin{figure}
\centerline{\mbox{\epsfxsize=9.5cm\epsffile{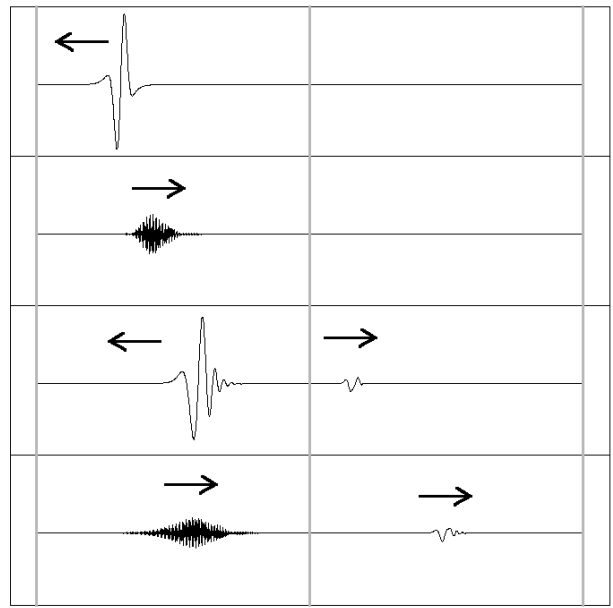}}}
\caption{
Four stages in the process of a wave packet reflecting off the white and 
black hole horizons. 
The black hole horizon is in the centre. 
These are the real parts of the wave packets 
(the imaginary look similar and were chosen to make the 
intial pulse purely positive norm). 
}
\label{evolution}
\end{figure}

This packet will now travel toward the white hole horizon. Here, there is
no packet on the superluminal side which travels away from the horizon.
The only possibility is that the packet be blue shifted at the white hole
horizon and come off as a mixture of the positive and negative large $k$
solutions. While one of these is positive norm, and the other is negative,
they are travelling together toward the black hole horizon. Here they
again emit a negative norm packet into the superluminal region, although 
if the phases were just right, the two could cancel and emit nothing into 
the superluminal side. Assuming that this does not happen, or that the 
conversion at the white hole is so small so as not to create any mixture, 
the subluminal packet's norm increases once again. After many such back and
forth reflections, the norm can grow arbitrarily large. One has an instability.
In Fig.\ \ref{evolution} we have four steps in this process. The black hole
horizon occurs in the center of the diagram, while the white hole horizon is at
the left and right edges (we assume periodic coordinates). The low frequency
intial wave packet travells toward the white hole horizon, and reflects as a
high frequency wave, dragged by the fluid. 
The third window is after the reflection from the back hole, 
and the fourth after the refection again from the white hole.

For small $\omega$ the amount of mixing of positive and negative norm modes 
at each of the horizons is governed by a thermal Bose-Einstein factor 
\bea
|\beta|^2\propto\frac{1}{\exp(\omega/T)-1}
\,,
\ea
with an effective temperature $T$ proportional to the effective surface 
gravity, i.e., the rate of change of the velocity or fluid flow across 
the horizon.
If one makes the white hole horizon have a very low effective temperature, 
one can minimize the creation of large $k$
negative norm solutions on ``reflection" of the packet from this horizon.
This means that the packet on the subluminal side will remain positive
norm, if it started thus. 

\begin{figure}
\centerline{\mbox{\epsfxsize=9.5cm\epsffile{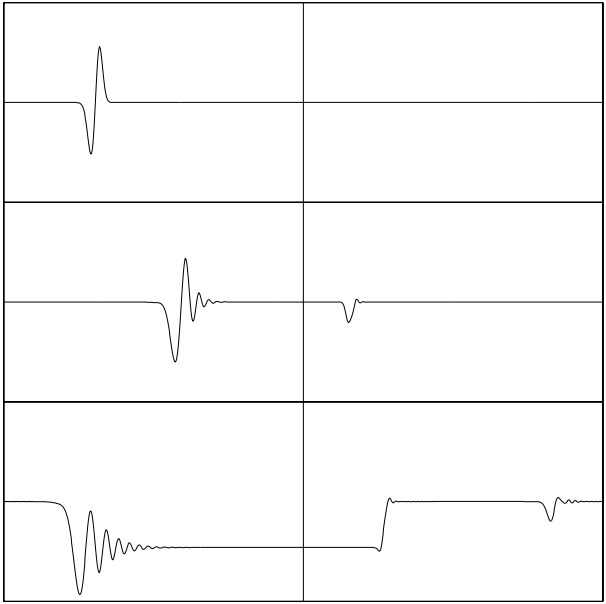}}}
\caption{
The reflection of a pulse producing low frequency instability. 
Plotted are three steps in the evolution of a pulse (top box) after one 
(middle box) and two reflections from the black hole horizon in the center. 
Note that in each case the pulse on the left (in the subluminal region) is 
travelling to the left while the ones on the right 
(in the superluminal region)  travel to the right. 
After another reflection from the black hole horizon, the pulse grows 
linearly and after still another, quadratically with distance from the black
hole horizon. 
}
\label{evolstep}
\end{figure}

One of the consequences of this instability is that one expects it to be 
worst for very small values of $\omega$. Since the norm mixing at
the black hole horizon is proportional to a Bose-Einstein factor, this
diverges for small values of $\omega$ 
\bea
{1\over \exp(\omega/T)-1}
\stackrel{\omega\downarrow0}{\longrightarrow}
{T\over\omega}
\,.
\eea
One would thus expect, and numerical simulations confirm, that after a few
back and forth bounces off the black hole horizon, the lowest wavenumber
mode of the wave (i.e., constant in space) would be amplified at each bounce.
Numerical simulation shows that after a couple of bounces, at each reflection
from the black hole horizon, a constant "step function like" pulse of ever
increasing amplitude is emitted by the black hole horizon into the subluminal
side, cf.\ Fig.\ \ref{evolstep}.
After the next reflection the spatial groth is linear, then quadratic etc.

One can, following  Corley and Jacobson \cite{laser}, make a very similar 
analysis in the case in which the dispersion relation is superluminal
\bea
F(k)=c^2\sqrt{k^2+a^2k^4}
\,,
\ea 
which is exactly the situation considered in Section \ref{surface}.
In that case one should consider a superluminal region between a 
black hole and a white hole horizon.  
In this case it is the wave packet bouncing between these two horizons on 
the superluminal side which creates the instability. 
Taking positive $\omega$ modes but with an intial negative norm wavepacket, 
we find that every time the packet hits the black hole horizon it emits a
positive norm packet into the subluminal region beyond that horizon
(which is again exactly the classical counterpart of Hawking radiation), 
thus increasing the size of the negative norm packet in the superluminal
region, while every time it bounces off the white hole horizon, 
it simply bounces off, possibly mixing in some
high frequency positive norm packet as well (depending again on the
effective temperature of the horizon).

Since with gravity waves, we can choose whether to have a subluminal or
superluminal dispersion relation, by choosing the depth and surface
tension of the liquid, one can test both sets of predictions. Furthermore,
since this norm mixing property of the horizon is intimately related to
the quantum emission of radiation by the horizon, one can, with purely
classical gravity waves, test the predictions about the thermal behaviour
of the various types of horizon.

The transition between short and long gravity waves can be described
by the height-dependent dispersion relation for a fluid at rest \cite{landau}
\bea
\omega^2=gk\tanh(kh_{\rm B})= 
gh_{\rm B}k^2\left(1-\frac{h^2_{\rm B}}{3}k^2\right)
+{\cal O}(k^6)
\,.
\ea
This should be contrasted with the dispersion of the surface tension waves
as derived in Eq.\ (\ref{dispersion}) of Section \ref{surface} 
(again for a fluid at rest)
\bea
\nonumber
\omega^2=gh_{\rm B}k^2\left(1+a^2k^2\right)
\,.
\ea
Consequently, one obtains a superluminal dispersion relation for 
$a \approx h_{\rm B}$ which implies $h_{\rm B}\approx2\;\rm mm$ for mercury.
For $a \ll h_{\rm B}$, on the other hand, the dispersion relation possesses 
a large subluminal region (short gravity waves) before it becomes
superluminal (surface tension waves) again.

\begin{figure}
\centerline{\mbox{\epsfxsize=6cm\epsffile{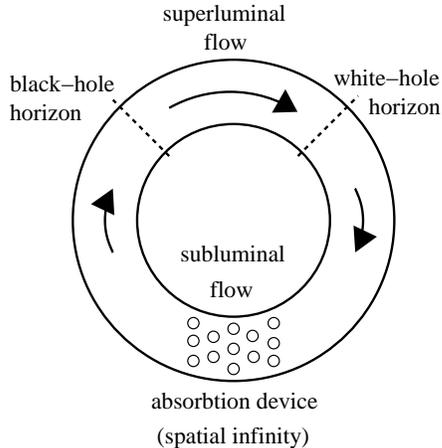}}}
\caption{
Sketch of a possible experimental set-up. A wavepacket is 
bouncing back and forth in the superluminal region between the
inner (white hole) and the outer (black hole) horizon,
emitting a piece of ``Hawking radiation'' at the black hole horizon
and thereby increasing its amplitude.
}
\label{ring}
\end{figure}

A possible experimental set-up for observing this instability is sketched
in Fig.\ \ref{ring}. 
The fluid is flowing within a circular basin with varying depth such that
the liquid's flow velocity as well as the speed of the gravity waves depends
on the position.
As the depth of the basin decreases the velocity of the fluid increases and
in the same time the speed of the gravity waves decreases.

In this way one may construct a region of superluminal fluid velocity
confined between the inner and the outer horizon.
As mentioned above, the gradient of the basin's depth 
(determining the the surface gravity) should be small at the inner 
(white hole) horizon whereas it should be as large as possible
-- without violating the assumption in Section \ref{model}, e.g.\ by 
generating a breakdown of the laminar flow -- at the outer (black hole)
horizon in order to observe the instability described above.

On the other hand, one also has to make sure that the wave packets emitted
by mode conversion into the other  region do not travel around and
disturb the system when they come back through the other horizon. 
For real black holes they escape to infinity. 
In our system, however, this is not the case and thus one has to install an 
additional absorbing device, which effectively models spatial infinity -- 
in order to prevent this interference.   

According to the remarks in Section \ref{viscosity} it appears to be
reasonable to move the bottom (e.g.\ made of rubber) of the fluid or
exert a magnetic force on it.
In practice one might generate a flow with a slowly increasing velocity
until the two horizons are formed.
In this case one should observe an instability of the laminar and smooth flow
profile exactly at this threshold velocity. 
Of course one has to ensure that there is no other instability at or before
this thershold which might spoil the experiment. 

\section{Discussion}\label{disc}

In view of the main advantage of the gravity wave analogs -- the possibility
of tuning the velocity of the wave propagation rather independently -- one 
would expect that it is possible to simulate black and white holes in the
laboratory and to study their instabilities experimentally. 
As one perhaps reasonable hierarchy of the different dimensions involved
$\delta h \ll h_{\rm B} \ll \lambda$
one could imagine waves of about 1~mm amplitude in a 1~cm deep basin with the 
wavelength being circa 10~cm and the characteristic size of the black or white
hole analog (its Schwarzschild radius) approximately 1~m.
The velocity of those waves -- and hence also that of the fluid -- in this 
case would be about $0.3\,{\rm ms}^{-1}$ and should be realizable.   
Aiming for the incorporation of surface tension waves for large wavenumbers
it might be suitable to divide the above suggested length scales by a factor 
of about 10.
Here we encounter another advantage of the gravity wave analogs:
In contrast to sound waves, for example, the amplitude of the gravity waves 
can be measured directly (as a length) and with a very high accuracy 
(e.g.\ via interferometry using the reflection at the fluid's surface).  

However, all the suggestions above have to be reconsidered when aiming for an 
experimental verification of the Hawking effect \cite{hawking}.
In order to detect this quantum instability the temperature of the fluid 
should be as small as possible -- whereas the fluid's velocity and its 
gradient have to assume their maximal feasible values.
The experimental verification of the Hawking effect seems to be on or even 
still beyond \cite{pessi} the edge of our present experimental capabilities 
-- for the dielectric/optical as well as for the sonic/acoustic black 
hole analogs (Dumb holes) and probably even more so for the gravity wave 
analogs.

\section*{Acknowledgement}

This work was supported by the Alexander von Humboldt foundation,
the Canadian Institute for Advanced Research, and the NSERC. 

\addcontentsline{toc}{section}{References}

\end{document}